# Generic solution for hydrogen-antihydrogen interactions

G. Van Hooydonk, Ghent University, Faculty of Sciences, Krijgslaan 281 S30, B-9000 Ghent (Belgium)

**Abstract**. An internal inconsistency with mutually exclusive Hamiltonians and the corresponding mutually exclusive states HH and H-antiH suffices to doubt the usefulness of *ab initio* calculations for H-antiH based on 1927 Heitler-London theory. Removing this inconsistency in a generic way invalidates this theory but the compensation is that problems with antihydrogen are removed.

Pacs: 34.10.+x, 34.90.+q, 36.10.-k

Claims that mass-production of $\underline{H}$ *seems* possible [1] boosted work on H-$\underline{H}$ interactions [2]. A charge-inverted Hamiltonian is needed to describe H$\underline{H}$ with *ab initio* procedures [2]. All these authors persistently overlook *the generic bonding character of this charge-inverted Hamiltonian*, as is demonstrated below. A result with a *generically bonding Hamiltonian* is easily predicted without calculation: *it must, by definition, always lead to a bound 4-particle system*. Since H$\underline{H}$ is considered as unstable [2], these methods are erratic when a charge-inverted Hamiltonian is proved attractive. This *historical* error is about *the mutually exclusive character of 2 (quantum) states* for HH and H$\underline{H}$ or their Hamiltonians. Our proof starts with the well-known non-relativistic 10 term Heitler-London (HL) Hamiltonian for molecular hydrogen $H_2$ (=HH) [3] abbreviated as

$$H_+ = H_0 + \Delta H \qquad (1)$$

where $H_0$ is the collection of the 6 intra-atomic terms and $\Delta H$ of the 4 inter-atomic terms. This Hamiltonian is also valid for $\underline{HH}$, since HH and $\underline{HH}$ are charge-symmetrical. HL showed that, for molecular hydrogen HH=$H_2$, the eigenvalues deriving from (1) are

$$E = E_0 \pm \beta \qquad (2)$$

where $E_0$ is the eigenvalue for atomic asymptote H+H and $\beta$ the eigenvalue for the interaction, i.e. the resonance or exchange integral of HL-theory [3]. But with (1), *β is also the eigenvalue of ΔH*, since $E_0$ is the eigenvalue of $H_0$ for the asymptote. HL-result (2) explains the observed splitting of states (bonding and anti-bonding) in molecular hydrogen but gave an *inaccurate* PEC (potential energy curve) for the bound singlet state. Despite this, HL-theory is considered as a triumph for wave mechanics, since exchange forces are absent in classical physics. With (2), the 2 quantum states for neutral 4-particle system, molecular hydrogen, are *mutually exclusive*, as expected.
For charge-asymmetrical systems H$\underline{H}$ and $\underline{H}$H, a charge-inverted Hamiltonian $H_-$ appears [2]. With (1) as reference, this is given by

$$H_- = H_0 - \Delta H \qquad (3)$$

the only valid Hamiltonian for studies on H<u>H</u>-systems [2]. As a result, all *charge-symmetrical* (HH and <u>HH</u>) and *charge-asymmetrical* systems (H<u>H</u> and <u>H</u>H) can now be described with only one algebraic Hamiltonian, having a built-in parity operator

$$H_\pm = H_0 \pm \Delta H \qquad (4)$$

*which, by definition, typifies Hamiltonians (1) and (3) as mutually exclusive too. This means literally that only one of the two Hamiltonians (1) or (3) can be bonding and lead to a stable 4-particle system, the other is doomed to be anti-bonding and must lead to a repulsive system. The real surprise with (4) however is that it is easily verified, without any calculation, that its two eigenvalues must be of type*

$$E_\pm = E'_0 \pm \beta' \qquad (5)$$

*and, therefore, are at least formally identical (degenerate) with those of HL-theory (2). Only, the two eigenvalues in (5) separate by virtue of intra-atomic charge-inversion, whereas those of HL-theory (2) separate by virtue of lepton spin-inversion.* Analytical details of the two approaches [4] show that, even if asymptotes $E_0$ in (2) and $E'_0$ in (5) and corresponding bond energies β and β' were different, this has no consequences to decide about the *generic character* of Hamiltonians (1) and (3). In fact, asymptotic freedom must be allowed anyhow for 4-particle system molecular hydrogen [4].

Before deciding which option is the better to explain a chemical bond, it is easily verified that, because of this formal degeneracy, charge-symmetrical state HH obeying (1) and charge-asymmetrical state H<u>H</u> obeying (3) are *two mutually exclusive states allowed for the 4-particle quantum system called molecular hydrogen*. Exactly this formal degeneracy of (2) and (5) leads to internal inconsistencies in studies like [2].

Choosing between the 2 formally degenerate options proves difficult, although it is known for long that charge- and spin-operators are identical, apart from a factor 2, which is consistent with the formal degeneracy of (2) and (5) also [4]. To reach a conclusion, the physics and chemistry community decided a long time ago to impose a rather drastic solution: the exclusion of (5) as a real solution by simply forbidding (5) in the natural world.

At first sight, this is understandable, since one is indeed tempted to characterize Hamiltonian (1) as *bonding*, since in HL-theory (2) only the charge-symmetrical Hamiltonian (1) leads to the stable bound state for molecular hydrogen. By exclusion, this automatically would not only classify Hamiltonian (3) as anti-bonding but also the quantum state H<u>H</u>, it represents. Interpreted with HL-theory and using (4), H<u>H</u> would be always be classified as a repulsive state.

Simple arguments prove that this drastic conclusion, i.e. forbidding (5), is wrong [4], which would invalidate the use of the Hamiltonian (1) of HL-theory. If so, the only generic solution possible through the mutually exclusive constraint, is that the bound state of molecular hydrogen is charge-asymmetrical H<u>H</u>, whereas its repulsive state is charge-symmetrical HH. This generic solution would immediately solve problems surrounding H̲, as argued before [4] but at the same time, invalidate the Hamiltonian in HL-theory as well as in its many successors.

For annihilation of charge-asymmetrical system HH̲ [2] to take place at short range, either asymptote $E_0$ or $E'_0$ must be reached from the attractive side, i.e. with (3), which proves the repulsive nature of (1) by exclusion. This proves (3) is attractive, since repulsive systems can never annihilate. For molecular hydrogen, annihilation does not actually occur, since there are too many repulsive terms in both (1) and (3), including the kinetic energy terms [4].

Also, there is abundant evidence [4] that the decisive parameter to describe the total energy of 4-particle systems (chemical bonds) is the inter-baryon separation, r(pp) in (1) and r(pp̲) in (3). It is evident that in HL Hamiltonian (1) this is repulsive $+e^2/r(pp)$, whereas it is attractive $-e^2/r(pp̲)$ in (3). *Also this proves that an attractive Hamiltonian for a 4-particle system can only be a charge-inverted Hamiltonian (3)*. If this logical consequence of (4) would be denied, even the universal Coulomb law should be invalidated, since this states that, by definition, like charges or charge-symmetrical systems (++; --) or (1) always repel, while unlike charges, charge-asymmetrical or charge-conjugated systems (+-; -+) or (3) always attract. This would indeed be the absurd result of denying the reality and the real meaning of (4). Therefore, HL-theory, based upon (1), clearly defies, contradicts and simply violates the universal character of the Coulomb law.

Hence, the asymptote is approached from the repulsive side with (1), whereas only with (3), it is reached from the attractive side, which settles the problem and proves that (3) is the only Hamiltonian to have generic attractive character for a 4-particle system like a chemical bond. A byproduct of this elementary deduction is that all HL-based, i.e. all charge-symmetrical (1)-based quantum-chemical, procedures are affronted with computational difficulties, *since they all invariantly try to explain a stable state using repulsive HL Hamiltonian (1)*. After decades of intensive quantum-chemical research, the many computational difficulties are proved *de facto*. This inconsistency also explains (a) why, in practice, wave functions needed to invert the character of (1) from repulsive to attractive are *unnecessarily* complex and (b) why they all must have a built-in parity operator like in HL-theory *to remedy for this inversion of (1)*.

Failing to recognize the generic attractive character of (3) leads to inconsistencies. If, in reality, the bound state of molecular hydrogen were HH̲, due to (5), many problems in theoretical and experimental physics and chemistry on H̲ would disappear, as argued before [5].

Many will find it difficult to admit that Heitler and London were wrong, be it *involuntarily*, since, at the time, charge-inversion problems did not even exist. Still, one is confronted with experimental evidence for natural atom H̲ [5]. To make sense, claiming the presence of a natural charge-inverted state H̲ in a bound state of molecular hydrogen on the basis of (5), implies that a charge-inverted H-state or H̲ must also show in the line spectrum of natural species hydrogen, which is exactly what we proved [5]. Unfortunately, none of these results is properly referenced [6], not even in a broad forum like a review on H̲ [7].

More consequences, arguments and references in favor of this analysis, including the striking similarity between bonding explained with **C**-symmetry as in (5) and the quark model of particle physics are in a more elaborate paper [4].